# Electron attachment to Cytosine: The Role of Water

Pooja Verma[a], Debashree Ghosh[b] and Achintya Kumar Dutta[a*]

*[a]Department of Chemistry, Indian Institute of Technology Bombay, Powai, Mumbai 400076*

*[b]Indian Association for the Cultivation of Science, Kolkata.*

**Abstract:** We present an EOM-CCSD based QM/MM study on the electron attachment process to solvated cytosine,. The microhydration studies cannot capture the effect of bulk water environment on the electron attachment process, and one need to include a large number of water molecules in the calculation to get converged results. The electron attachment in the bulk solvated cytosine happened through a doorway mechanism, where the initial electron is localized on water. The electron subsequently gets transferred to cytosine by mixing of electronic and nuclear degrees of freedom, which happens at an ultrafast time scale. The bulk water environment stabilizes the cytosine bound anion by an extensive hydrogen-bonding network and enhances the electron transfer rate by manifold from that observed in the gas phase. The predicted adiabatic electron affinity and electron transfer rate obtained from our QM/MM calculations are consistent with the available experimental results.

*achintya@chem.iitb.ac.in

# 1. Introduction

Attachment of low energy electron plays a crucial role in the radiation damage pathways of genetic materials[1]. Water comprises around 70-80% by weight of the human body and plays a significant role in all the biological processes in living organisms, including radiation damage. A significant part of the high energy radiation (UV, X-ray, gamma radiation, ion-beam) is absorbed by liquid water surrounding the biomolecules, and it generates transiently excited water molecules ($H_2O^*$), water cation ($H_2O^{\bullet+}$), and secondary electrons (SE)[2–5]. The excited $H_2O^*$ then autoionizes to form $H_2O^{\bullet+}$ on a femtosecond timescale[6], and subsequently gets deprotonated to generate $OH^{\bullet}$ and $H_3O^+$. The $H_2O^{\bullet+}$ can also undergo dissociation to form $OH^{\bullet}$ and $H^{\bullet}$. All these free radicals can attack the DNA subunit and lead to structural distortion and bond breaking. However, secondary electrons (SE) are also a major source of radiation damage. The SE can possess a broad distribution of energies varying from < 5 eV to 30 eV. They subsequently lose energy via collision and vibrational relaxation to generate low energy electrons (LEE).

These LEE come in contact with water and undergo a multi-step solvation process. In the first step, they form quasi-free electron ($e^-_{qf}$), which has zero kinetic energy. The $e^-_{qf}$ in the next step gets converted into pre-hydrated electrons ($e^-_{pre}$), which are rapidly solvated by the surrounding water molecules within a picosecond to form the "solvated electron" ($e^-_{aq}$). The radiation chemistry of low energy electrons depends on various factors such as energy, the extent of solvation and surrounding reactive species. The various stages of water ionization and the mechanism of formation of reactive radicals and secondary electrons are shown in Figure 1. Over the last decade, numerous experimental studies have tried to unravel the role of LEE in the radiation damage pathway. Sanche and co-workers have presented one of the first study[7] on the DNA strand cleavage induced by LEEs using electrophoresis. This has inspired a variety of mechanistic studies[8–10] on DNA sub-units in the gas phase or condensed states. These studies proposed the formation of a transient negative state due to electron attachment to DNA sub-unit in the gas phase, which then leads to fragmentation via dissociative electron attachment process. Similar fragmentation behaviour has been showed for solvated DNA sub-units using photo-electron imaging[11], LEE scattering[12], and Rydberg electron transfer spectroscopy[13].

In recent times, ultrafast transient spectroscopy[14,15] methods have provided much accurate mechanistic insight into the electron attachment to solvated nucleobases and nucleotides. These have highlighted a crucial role played by the excited state of the nucleobases in the electron attachment process. However, one is yet to achieve a proper understanding of the mechanism of electron attachment to DNA sub-unit in the presence of water. Theoretical simulations can be helpful in the interpretation of the experimental results. Consequently, a lot of theoretical studies on electron attachment to solvated nucleobases, base-pairs, and nucleotides are reported in the literature[16,17]. However, most of the reported theoretical results are based on the implicit solvation model[18,19]. An implicit solvation model like PCM can take care of the polarization effect induced by the solvent. However, it fails to account for the various explicit interactions[20], including Hydrogen bonding. These direct interactions play a major role in stabilizing the electron and anionic DNA subunits. Therefore, an explicit treatment of water molecules is necessary for getting quantitative accuracy in the simulation of solvated electrons. However, an explicit treatment of bulk water molecules in a quantum chemical method is not possible and restricts the number of water molecules included in the calculations to a very small number. The inclusion of a sufficient number of water molecules in the simulation for describing the bulk solvation requires a multilevel approach like QM/MM[21].

There are prior reports of the QM/MM based explicit solvation studies on the electron attachment to nucleobases existing in the literature[22–25]. However, all of them use the DFT method for the QM region. The standard DFT functionals perform poorly for anions due to the self-interaction problem[26]. However, the wave function-based methods are free from the problem of self-interaction. Unfortunately, the electron affinity in wave-function-based methods converges very slowly with respect to the size of the basis sets. Therefore, one needs to use a very large and flexible basis set with a sufficient number of diffuse functions to obtain reliable results, which makes the calculation very costly, sometimes even impossible to compute. Moreover, a large basis set with many diffuse functions increases the possibility of a variational collapse of reference anionic wave function in the $\Delta$ based methods. The direct energy difference methods are generally more stable with respect to the variational collapse and can calculate electron affinity corresponding to multiple states in a single calculation. Additionally, the direct energy-difference-based method gives information about the transition process, which is difficult to obtain in $\Delta$ based methods.

Among the various direct energy-difference-based methods available, the equation of motion coupled cluster (EOM-CC) method[27–30] is considered to be the most accurate and systematically improvable. The EOM-CC is generally used in the singles, and doubles approximation (EOM-CCSD), which scales as $O(N^6)$ power of the basis set and has a storage requirement, which scales as $O(N^4)$. Due to high computational cost and huge storage requirement, the use of a large basis set restricts the applicabity of canonical EOM-CCSD method beyond isolated nucleic acid base. However, newly developed Pair Natural Orbital (DLPNO)[31–39] based EOM-CCSD method[40–45] can handle electron attachment to large molecules on an extended basis set with systematically controllable accuracy. We have recently developed an EA-EOM-DLPNO-CCSD based QM/MM protocol to study electron attachment to solvated nucleobases[46].

It has been observed that in the gas phase calculation, the electron attachment leads to the formation of cytosine bound states in the GC base pair[47,48], and it is more stable than the corresponding thymine bound states in AT bases. Smith and Kohanoff[22], and Schaefer and co-workers[16] have shown that the initial attachment of electron in aqueous phase happens to the nucleobase part of nucleotide, and the nucleotide containing cytosine shows the lowest barrier for sugar-phosphate bond breaking. Therefore, it is important to understand the mechanism of electron attachment to cytosine in an aqueous environment. The work investigates the mechanism of electron attachment to solvated cytosine using high-level EOM-DLPNO-CCSD based QM/MM methods.

## 2. Computational Details:

The micro-solvated cytosine structures were optimized using RI-MP2/def2-SVP level of theory. We have considered 10 low lying isomers (Fig 2.), and optimized geometries have been provided in the supporting information. The single point electron affinities were calculated using EA-EOM-DLPNO-CCSD/aug-cc-pVTZ level of theory. Additional 5s5p4d diffuse functions[49] were added to the positive end of the dipole moment. RIJK approximation in ORCA has been used for Hartree-Fock calculations, and the NORMALPNO setting has been used.

To mimic the effect of bulk water, the cytosine was placed in a water box of 40Å length and solvated with 2527 water molecules. Classical molecular dynamics (MD) simulations were performed to generate the distribution of water molecules around cytosine molecule. Classical MD simulations were performed using the NAMD software package[50].A CHARMM compatible forcefield, parameterized with VMD Force Field Toolkit (fftk) has been used for the cytosine, and the TIP3P model has been used for the water molecules. The forcefield files for the cytosine can be downloaded from our website[51].

In the first step, a minimization was performed, keeping the cytosine fixed with a non-bonded interaction cut off of 12 Å. Subsequently, the system is slowly heated up to 300 K at constant volume using Langevin thermostat. In the next step, the system is equilibrated for 1 ns. The constant temperature of 300 K and the constant pressure of 1 atmosphere have been kept using a Parrinello-Rahman barostat. The periodic boundary condition and particle-mesh Ewald technique (PME) has been used for the equilibration run. Constraints were put on the bonds involving hydrogens using the LINCS algorithm. Subsequently, a 10 ns production run was carried out in constant pressure (1 bar) and constant temperature (300 K) condition with Nosé-Hoover Langevin piston pressure control. A time integration step of 2 fs time step has been used for the equilibration and production run.

Ten equally spaced snapshots were generated at an interval of 1 ns from the classical MD trajectory, which are used for the subsequent QM/MM molecular dynamics simulations.Cytosine is considered as the QM region for the QM/MM MD run and is treated at the BP86/def2-SVP level of theory. The water molecules are treated at the MM level using the same forcefield as that of the classical MD simulations. QM/MM MD simulations were performed for 2 ps on each of the 10 snapshots from the classical MD. A time step of 0.5 fs has been used. The calculations were performed using a box length of 40 Å. The periodic boundary condition and non-bonded cutoffs were not used for the QM/MM calculations. After discarding the first 1 ps of trajectory, 40 snapshots were generated at an equal interval of 25 fs from the last 1 ps of each trajectory. The single point QM/MM calculations for all 400 snapshots (taken 40 from each trajectory) were performed using ORCA, treating cytosine with EA-EOM-DLPNO-CCSD/aug-cc-pVDZ method with NORMALPNO setting, and TIP3P model for water. An additional 5s5p4d diffused functions were placed on the carbonyl oxygen of cytosine.

Total 55 snapshots for the calculation of adiabatic electron affinities (AEA) were generated from the 400 QM/MM snapshots using an energy based pre-screening scheme with a cutoff value of 0.09 eV. These 55 snapshots were separately optimized for the neutral as well as anion geometry using QM/MM method. The details of the protocol for the calculation of the adiabatic electron affinity of the solvated nucleobases can be found in reference[46].

The ORCA quantum chemistry package is used for all QM and QM/MM calculations.

## 3. Results and Discussion:

### Micro-solvation:

The monohydrated cytosine represents a simple model system to investigate the role of water in the process of electron attachment to the cytosine. Table 1 presents dipole-moment, vertical electron affinity (VEA), vertical detachment energy (VDE), and adiabatic electron affinity (AEA) of bare and all the ten isomers of monohydrated cytosine in EA-EOM-DLPNO-CCSD/CBS level of theory. The addition of an electron at the neutral geometry leads to a dipole bound state in bare cytosine where the additional electron is located away from the nuclear framework. Similar dipole bound states are observed in some of the monohydrated cytosines. All the monohydrated isomers show a dipole moment smaller than the dipole moment of bare cytosine. Consequently, they have a VEA value, which is less than the bare cytosine. These dipole bound states can act as a doorway for the formation of a valence bound state (See Figure 3). The valence bound state is not bound at neutral geometry and can be observed only at anion geometry. The transition between the two anionic states is very feebly optically allowed with an oscillator strength of only 0.004. However, the molecular vibrations can also result in the inter-conversion of the two states (valence and dipole-bound state of the anion). To understand the mechanism of interconversion, we have plotted the adiabatic potential energy surface (PES) of the ground and first excited state of the anion along a linear transit from the dipole-bound to valence-bound geometry for the lowest energy isomer a (Figure 4). Intermediate geometries along the linear transit can be derived using the following expression:

$$R_{int} = (1-\lambda)R_{DB} + \lambda R_{VB} \qquad (1)$$

Where $R_{DB}$ is the parameter (bond length, bond angle, and dihedral angle) for the dipole-bound geometry, and $R_{VB}$ is the parameter corresponding to the geometry of the valence bound state. The $\lambda$ is the linear transition parameter, which varies from 0 to 1. The $\lambda = 0$ leads to the dipole-bound geometry, and $\lambda = 1$ leads to the valence-bound geometry.

The potential energy surface shows two minima corresponding to dipole and valence-bond state, respectively. The ground and first excited state of the anion show avoided crossing where the nature of the electronic wave function changes rapidly. This increases the derivative coupling term and leads to the breakdown of the Born-Oppenheimer approximation. One can treat such cases in terms of diabatic potential energy surfaces. The diabatic surfaces cross each other along the path connecting the two minima, and the dominant part of the coupling term is shifted to electronic Hamiltonian. Therefore, the electronic coupling between the two diabatic states replaces the coupling between nuclear and electronic degrees of freedom. The definition of the diabatic states are not unique. However, the valence and dipole-bound nature of the state give us an obvious choice for the diabatic basis[52,53].

We have generated the two diabatic states by fitting a harmonic potential to the valence-bound and dipole-bound part of the ground state adiabatic PES of mono-hydrated cytosine anion. The diabatic potential energy surface of isomer a is presented in Figure 5. A two-state avoided crossing model[54] potential was used to calculate the coupling between the two diabatic states as follows:

$$V = \begin{pmatrix} V_1 & W \\ W & V_2 \end{pmatrix} \qquad (2)$$

where the diagonal elements $V_1$ and $V_2$ are harmonic potentials in the coordinate $\lambda$ as per the following equation, keeping off-diagonal elements constant. The potential varies with $\lambda$ as per the equation:

$$V_i = \frac{1}{2}\omega_i(\lambda - \lambda_i^0)^2 + V_i^0 \quad (3)$$

The coupling element for isomer a is found to be 7.9 meV, and anionic states of cytosine are in the threshold of weak coupling limit. The rate of transition of an electron from dipole-bound state to valence-bound state in the weak-coupling limit is estimated using Marcus theory[55] as:

$$k = \frac{2\pi}{\hbar}|W|^2 \sqrt{\frac{1}{4\pi k_B T \lambda_R}} e^{\frac{(\lambda_R + \triangle G^0)^2}{4\lambda_R k_B T}} \qquad (4)$$

where $\lambda$, W and $\triangle G^0$ represents the reorganization energy, coupling constant, and free energy change between the two states ($E_{VB} - E_{DB}$, ignoring the entropy contribution) respectively.

The calculated rate of transition from the dipole-bound to valence-bound state is $1.9 \times 10^7$ sec$^{-1}$ at room temperature for the isomer a. The rates are much slower than that observed for microhydrated uracil[56] or microhydrated GC base pair[57]. However, the microsolvation leads to at least two-fold increase in the rate of transition from the gas phase value of $1.6 \times 10^5$ sec$^{-1}$. However, isomer a is not bound adiabatically and will undergo auto-detachment. Therefore, the doorway mechanism in this case serves as the decay pathway of the dipole bound anion.

None of the microhydrated cytosine isomers is adiabatically bound. However, these isomers show a wide variation in the AEA values among themselves. The isomer b shows the highest AEA value of -0.07 eV among all the isomers. The same isomer shows the highest VDE of 0.96 eV, which is consistent with experimental photodetachment photoelectron spectroscopy studies.[58]

## Bulk-solvation:

### Vertical detachment energy of bulk solvated cytosine:

Although a microsolvated model can be important for an understanding of the mechanism of the electron attachment process, one needs to consider the effect of bulk water to simulate the experimentally observed electron attachment process. Schiedt *et al.*[58] have shown that the electron affinities of solvated cytosine increase linearly with the number of water molecules. The bulk-water environment can drastically increase the stability of the anion by an extensive hydrogen-bonding network. Figure 6 presents the instantaneous and average VDE calculated for 400 snapshots taken from QMMM dynamics. It shows an average value of 8.24 eV, which is much larger than that observed from the microsolvation study. Figure 7 presents the radial distribution function plot for (a) primary amine nitrogen of cytosine and hydrogen of water (b) cytosine oxygen and hydrogen of water for the solvated cytosine. In both cases, one can see a sharp peak around 2 Å, which indicates a short-range local structure of water is formed around the cytosine anion. This local water structure leads to the preferential stabilization of the cytosine anion through hydrogen bonding, which results in high VDE values.

To further understand the distance dependence of the stabilizing effect provided by water, we have studied the effect of the size of the water environment on the calculated VDE values. First 55 snapshots are selected out of the original 400 snapshots using an energy-based pre-screening scheme[59,60]. The average VDE of these 55 snapshots is 8.29 eV, which is in good agreement with the original average value of 8.24 eV. In the next step, we have increased the number of water molecules included in the EA-EOM-DLPNO-CCSD based QMMM calculations in a shell wise manner for these 55 snapshots. Cytosine has been treated using the EA-EOM-DLPNO-CCSD method, whereas the water molecules are treated using TIP3P forcefield. A shell length of 2.7 Å has been chosen following the work of Ghosh and co-workers[61]. Figure 8(a) shows that the VDE value increases with the inclusion of extra shell for each and every snapshot. The VDE of bare cytosine 0.07 eV. The average value of the VDE with the first shell of water is 2.63 eV. The VDE increases to 4.46 eV as we move from the first shell to the second shell. The VDE value further increases by 1.16 eV and 0.79 eV with the inclusion of third and fourth shells. The VDE value starts to reach convergence on reaching the 6th shell and the change in average VDE from 6th shell to 7th shell in 0.3 eV. The single-point QM/MM calculations with periodic boundary conditions are not currently possible with ORCA. Therefore, the QM/MM calculations performed in this paper are essentially performed on large water clusters. To calculate the VDE for cytosine anion in bulk water, we have extrapolated the shellwise average VDE to infinite limit (see Figure 7(b)) by using an exponential fit and obtained a VDE value of 8.95 eV for cytosine anion in bulk water.

The calculated VDE values, especially in wavefunction-based methods, are extremely sensitive to the size of the basis set used in the calculation. To investigate the basis set dependence of the VDE values, we have performed a further pre-screening with a tighter threshold value of 0.02 eV. It has resulted in 20 snapshots with an average value of 8.17 eV in the aug-cc-pVDZ basis set, which is in good agreement with the VDE value of the original 400 snapshots. The VDE value increases 8.34 and 8.41 on going to the aug-cc-pVTZ and aug-cc-pVQZ basis set. Extrapolation to a complete basis set limit gives a value of 8.46 eV.

Jordan and co-workers[62] have shown that the use of non-polarizable water leads to over stabilization of the anionic states of the solvated electron. In a standard, non-polarizable QM-MM calculations, the MM region can polarize the QM region, whereas the back polarization of the MM region due to the QM region is not taken into account. One can increase the size of the QM region to account for the polarization effect. We have taken the 20 pre-screened snapshots and calculated the VDE in EA-EOM-DLPNO-CCSD/aug-cc-pVDZ based QMMM method with the progressively increased size of the QM region. The 2 Å in Figure 9 means any water molecules whose one of the atoms falls within the 2 Å distance of any atoms of cytosine are entirely considered in the QM region. One can see that the VDE value decreases with an increase in the size of the QM region for every snapshot. The instantaneous average value decreases for almost every snapshot (Figure 9(a)) with an increase in the size of the QM region. This shows that lack of back polarization due to non-polarizable water model leads to the over stabilization of the solvated cytosine. One can extrapolate the changes in VDE values with respect to the size of the QM region to complete box size 40 Å, and it results in a VDE of 6.59 eV. Although the increase in the size of the QM region can take care of the short-range part of the polarization, it cannot incorporate the long-range part of the polarization. However, the effect of the long-range part of the polarization[63] is generally small.

The calculation of final average VDE of cytosine solvated in liquid water as defined in the protocol described ref[46]:

Final average VDE= Average VDE of 400 snapshots from the EA-EOM-DLPNO-CCSD/aug-cc-pVDZ based QM/MM calculation + basis set correction+ correction for the finite number of water molecules + correction for the size of the QM region=8.24 eV +0.29 eV +0.66 eV-1.58 eV= 7.61 eV

**Adiabatic electron affinity of bulk solvated cytosine:**

Figure 8 presents the instantaneous and average AEA for the for 55 snapshots calculated using EA-EOM-DLPNO-CCSD/aug-cc-pVDZ based QM/MM methods. The average AEA value was for 55 snapshots is 3.68 eV. The AEA value increases to 3.77 eV and 3.83 eV on going to aug-cc-pVTZ and aug-cc-pVQZ basis set respectively. The AEA value at the extrapolated complete basis set limit is 3.87 eV. The zero-point energy (ZPE) corrections were calculated by using a partial Hessian analysis of the cytosine moiety using BP86/def-SVP based QM/MM method. The correction is found to be small at 0.09 eV, which is consistent with the previously reported values[24,46] for solvated nucleobases. We have approximated correction due to the finite size water box and the size of the QM region to be the same as the VDE.

Final average AEA value = Average AEA of the 55 snapshots from the EA-EOM-DLPNO-CCSD based QM/MM calculation + basis set correction + correction for the finite number of water molecules + correction for the size of the QM region+ZPE correction =3.68 eV+0.19 eV+0.66 eV-1.58eV+0.09 eV=3.04 eV.

The results are in excellent agreement with the AEA value (3.0 eV) reported Abel and co-workers[64] from their QM/MM calculation on bulk solvated cytosine.

Although there is no direct experimental measurement of the AEA value of cytosine in bulk water is available, one can estimate the AEA from the experimental reduction potential of cytosine with respect to the standard hydrogen electrode (SHE)

$$C + e \rightarrow C^-$$

$$AEA = \Delta G_C - \Delta G_{C^-} = -\left(E_{SHE} - E_{C/C^-}\right)\frac{F}{N_A e}$$

where F is the Faraday constant (96485 J mol$^{-1}$ C$^{-1}$), N$_A$ the Avogadro number and e the elementary charge. The reported experimental value[65–67] for the reduction potential of cytosine ($E_{C/C^-}$) varies quite widely from one another and shows a range of -0.81 eV to -1.44 eV. Taking the value of E$_{SHE}$ to be -3.281V following Schelgel and co-workers[68], the experimental AEA is estimated to be in the range 2.84 eV to 3.47 eV. Our calculated AEA values are well within experimental range.

**Mechanism of electron attachment in bulk water:**

The initial bound anionic states formed by the attachment of electron to solvated nucleobases are localized on water (at 0 fs). Figure 11 shows the EA-EOM-DLPNO-CCSD natural orbitals corresponding to the anionic states localized on water and cytosine for the first few snapshots from the first QM/MM trajectory (taken after 1ns of classical MD run). We have previously shown that the initial water-bound states of nucleobases arise primarily due to the interaction of the electron with bulk water[56]. The anionic states localized on the cytosine starts to appear within 5.5 fs of the initial electron attachment. The cytosine-bound state first appears as an excited state of the anion, which is very weakly bound with vertical detachment energy of 0.011 eV. The anionic ground state is still localized on the water (0.328 eV), and the transition from water-bound anionic state to cytosine-bound anionic state is very weakly optically allowed with an oscillator strength of 0.002, similar to that observed for the mono-hydrated cytosine model. Within 6.5 fs, the cytosine-bound state becomes the ground state with VDE of 0.450 eV. The water-bound states appear as the first excited state, with an excitation energy of 0.339 eV. Figure 12 shows the time evolution of detachment energy for the first six lowest energy states. One can see the presence of an avoided crossing in the bulk solvated cytosine anion similar to that in the adiabatic potential energy surface of microhydrated cytosine. Figure 13 presents the time evolution of the lowest anionic state localized on water and the lowest anionic state localized on cytosine. We have performed a crude diabatization by visual inspection of the natural orbitals. The ground state of bulksolvated cytosine anion at the neutral geometry is water-bound and the water-bound state acts as a doorway for the formation of the cytosine bound anion. The VDE corresponding to the cytosine bound state increases more steeply with time and gets converted into ground state of the anion with few femtoseconds. Such doorway mechanism will be competitive to resonance induced electron attachment to nucleobases in water[69–71].

Therefore, initial water-bound state tends to serve as a doorway for the electron attachment, which subsequently gets transferred to the cytosine bound state through the mixing of electronic and nuclear freedom. The reduction of nucleobases in water is experimentally observed [15] between $0.6 \times 10^{12}$ M sec$^{-1}$ to $5 \times 10^{12}$ M sec$^{-1}$. A plausible mechanism thus can be inferred from our theoretical simulations:

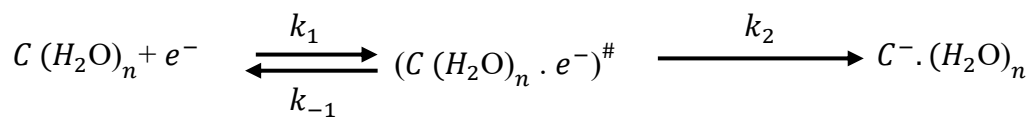

The first step includes the reaction of pre-solvated electron with cytosine to form an activated complex $(C\,(H_2O)_n\,.\,e^-)^{\#}$, where the extra electron is localized over water. In the next step, electron transfer happens from water to cytosine and the complex subsequently gets converted into solvated cytosine anion. Now, it is already known that the complete solvation of pre-hydrated electron takes place at picosecond timescale. Therefore, to sustain the experimentally observed rate of reduction of nucleobases in water, the electron transfer from water to cytosine should happen at a picosecond time scale.

One can estimate the rate of transfer of electron from the initial doorway states localized on water to the final anionic states localized on cytosine using a two-state model, where the reorganization energy and free energy of activation are calculated from the 55 snapshots used for AEA calculation and coupling constant is approximated to be same as in microhydration (w=7.9 meV). The calculated rate of electron transfer in bulk water is $0.6 \times 10^{12}$ at room temperature, which is in good agreement with the experimentally observed rate. The calculated rate is 5-fold larger from that observed for microsolvated states which highlight the crucial role played by the bulk water in electron attachment process. The rate of electron transfer from the initial water localized state to the final nucleobase localized state in bulk solvated cytosine is almost half of that observed for cytosine in bulk solvated GC base pair[57]. This indicates the cooperativity of the base-pairing with the solvation in the electron attachment process.

## 4. Conclusions:

We have studied the role of water in the electron attachment process to cytosine. The microsolvated structure shows a doorway mechanism of electron attachment where the initial electron attachment leads to a dipole-bound state, which acts as a doorway for electron capture.

The dipole-bound state subsequently gets converted into a valence-bound. The ground and the first excited state of the anion show an avoided crossing, and the transfer of electron takes place through a non-adiabatic process. The presence of even a single water molecule increases the rate of transfer of an electron from the initial dipole-bound state to the final valence-bound state. The rate of electron transfer in microhydrated cytosine is two-order of magnitude larger than that observed for isolated cytosine. The anionic states in bulk solvated cytosine show similar avoided crossings. The initial doorway states are localized on water and subsequently get transferred to cytosine bound states through a non-adiabatic process similar to that observed for the microsolvated structure. The bulk water stabilizes the cytosine anion by an extensive network of hydrogen bonding and enhances the rate of transfer of the electron from the anionic states localized on the water to the anionic states localized on cytosine. The rate of transfer of electron in bulk solvated cytosine is 5-fold larger than the microhydrated one.

The electron affinity of bulk solvated cytosine is found to be much larger than generally observed for microsolvated model, and the final electron affinity value in the used QM/MM method has been found to be sensitive to the size of the water environment, size of the QM region and the basis set used for the QM/MM calculations. The calculated rate of electron transfer and AEA of cytosine in bulk water shows excellent agreement with available experimental results.

The electron attachment can also lead to bond breaking in nucleotide and larger model systems, which will be competitive to the formation of the stable anion. It will be important to estimate the relative rates of these two processes in bulk water to understand the broad picture of the secondary radiation damage pathway of genetic materials. The work is in progress towards that direction.

## Supporting Information

The Supporting Information is available.

Cartesian coordinates of the snapshots used for VDE and AEA calculations, the VDE values for the pre-screened snapshot, the individual VDE values for shell wise convergence of water environment, the individual VDE values for convergence with respect to the size of the QM region, individual energy of the neutral and anionic state for AEA calculation are provided in the supporting information.

## Acknowledgment

The authors acknowledge the support from the IIT Bombay, IIT Bombay Seed Grant project, DST-Inspire Faculty Fellowship for financial support, IIT Bombay super computational facility and C-DAC Supercomputing resources (PARAM Yuva-II) for computational time.

## Conflict of interest

The authors declare no competing financial interest.

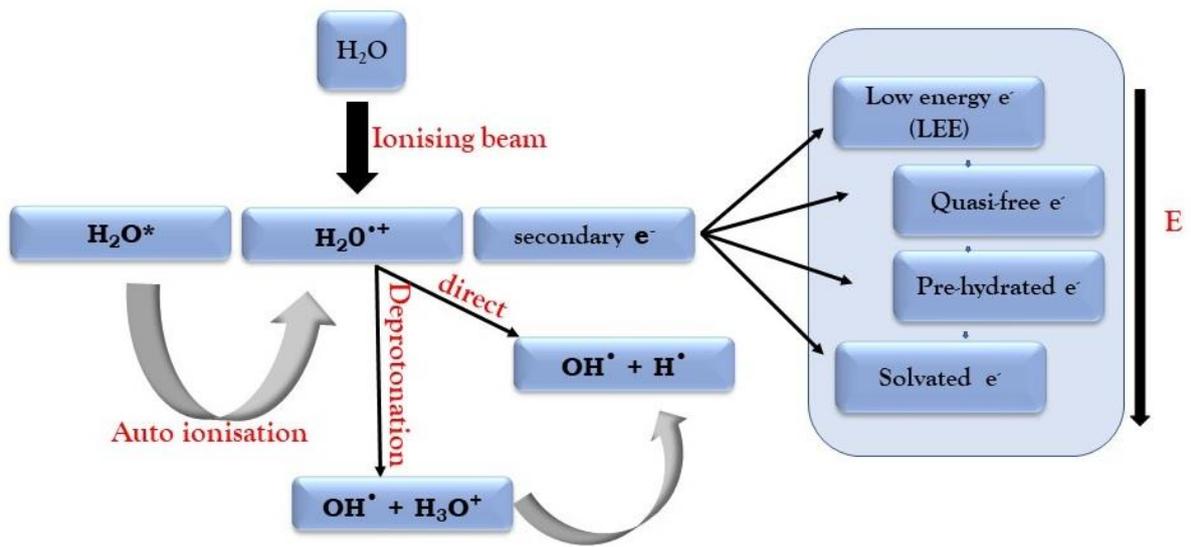

*Figure 1: Water ionisation resulting into ions, radicals and free electrons.*

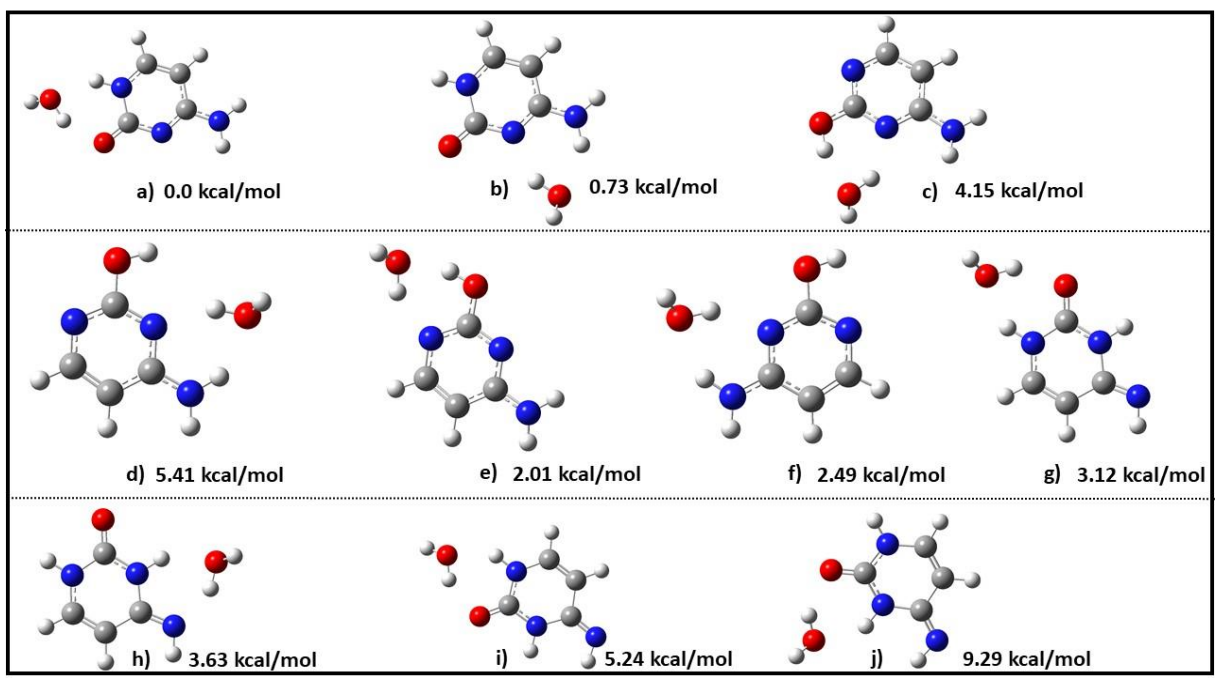

*Figure 2: The ten lowest lying monohydrated cytosine isomers. The relative energy (in kcal/mol) with respect to the lowest energy isomer at B3LYP/6-31++G\*\* level of theory is provided.*

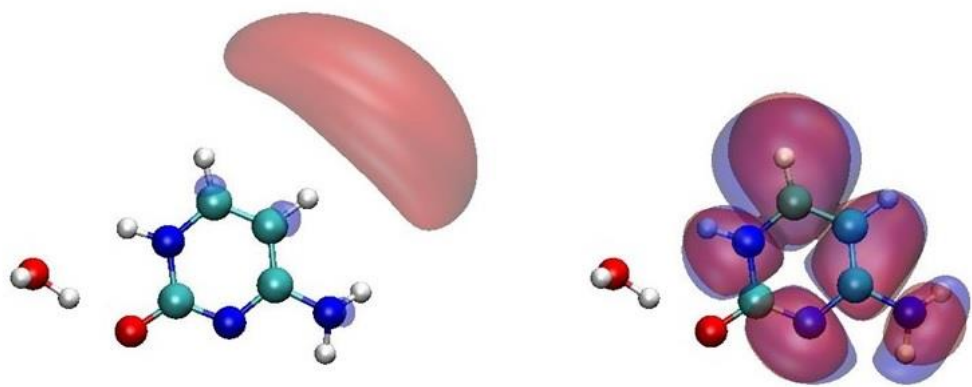

*Figure 3: Dipole and Valence bound state of monohydrated cytosine.*

**Table 1: The VEA, VDE and AEA values of bare and mono-hydrated GC at EA-EOM-DLPNO-CCSD/CBS level of theory. The dipole moment values are calculated for at the Hartree-Fock level.**

| Molecule | Dipole Moment (Debye) | VEA (eV) | VDE (eV) | AEA (eV) |
|---|---|---|---|---|
| Bare Cytosine | 6.89 | 0.08 | 0.07 | -0.08 |
| Micro-hydrated isomer a | 6.52 | 0.04 | 0.83 | -0.13 |
| Micro-hydrated isomer b | 7.03 | 0.06 | 0.96 | -0.07 |
| Micro-hydrated isomer c | 5.65 | 0.04 | 0.33 | -0.46 |
| Micro-hydrated isomer d | 3.63 | 0.0 | 0.37 | -0.42 |
| Micro-hydrated isomer e | 4.70 | 0.01 | 0.36 | -0.54 |
| Micro-hydrated isomer f | 2.95 | 0.0 | 0.34 | -0.50 |
| Micro-hydrated isomer g | 4.06 | 0.06 | 0.66 | -0.29 |
| Micro-hydrated isomer h | 6.49 | 0.04 | 0.80 | -0.16 |
| Micro-hydrated isomer i | 2.10 | 0.00 | 0.61 | -0.37 |
| Micro-hydrated isomer j | 3.91 | 0.01 | 0.62 | -0.37 |

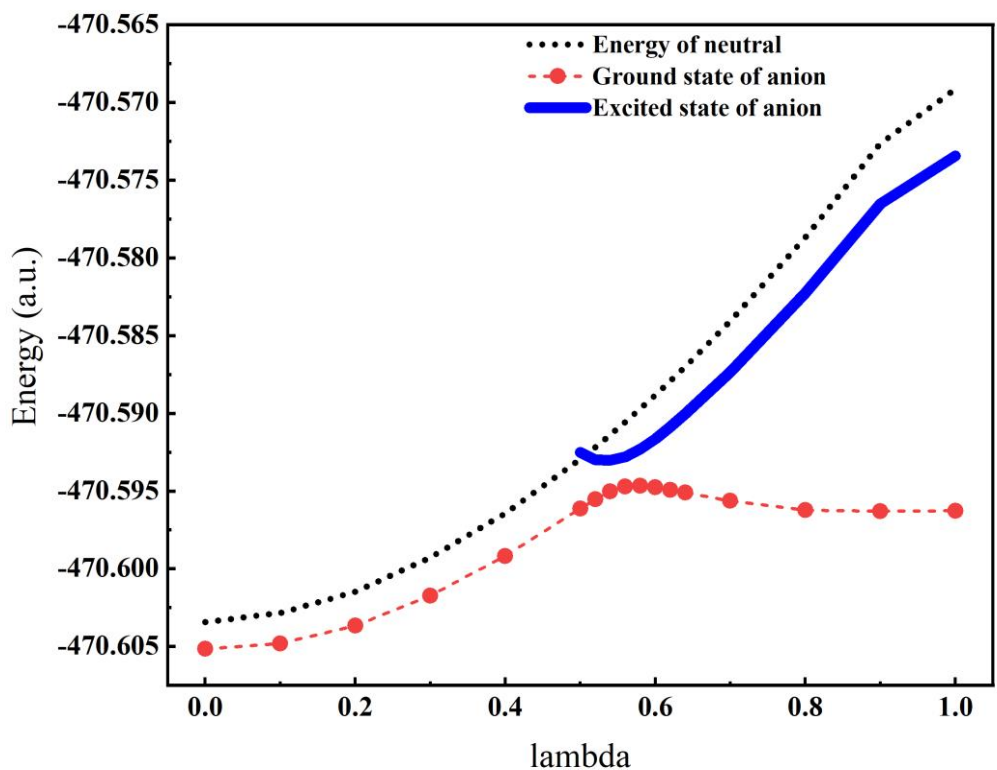

**Figure 4:** *Adiabatic potential Energy Surface (PES) of ground and first excited state of cytosine anion at EA-EOM-DLPNO-CCSD/aug-cc-pVTZ.*

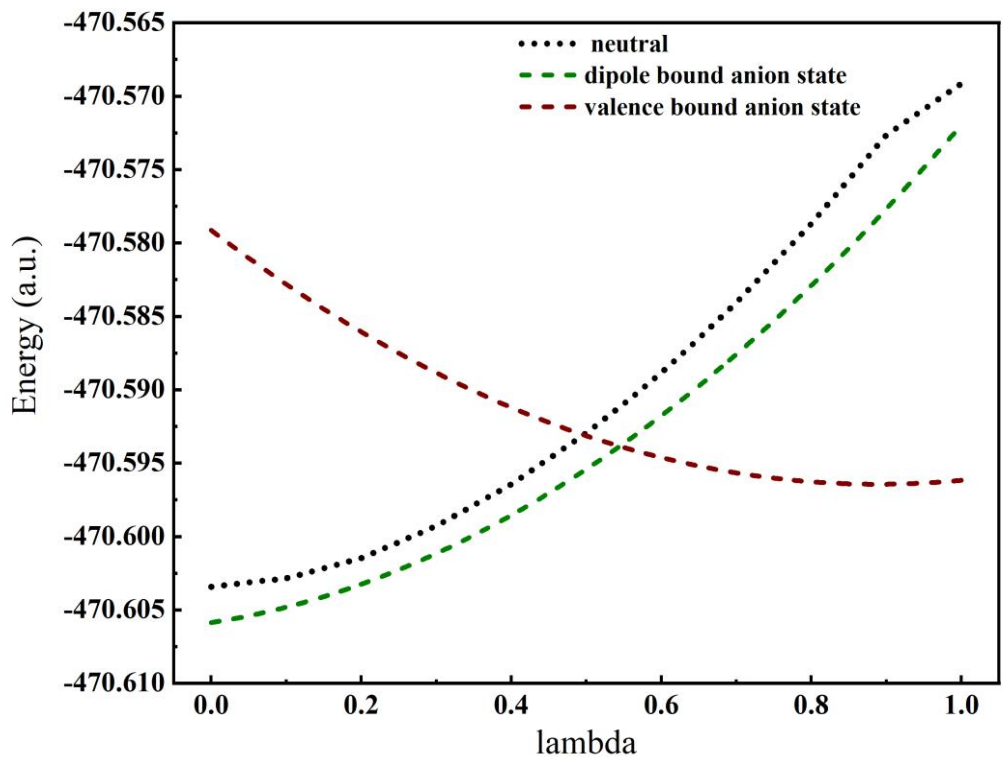

***Figure 5: Diabatic potential energy surface corresponding to valence and dipole bound state of microhydrated cytosine isomer a.***

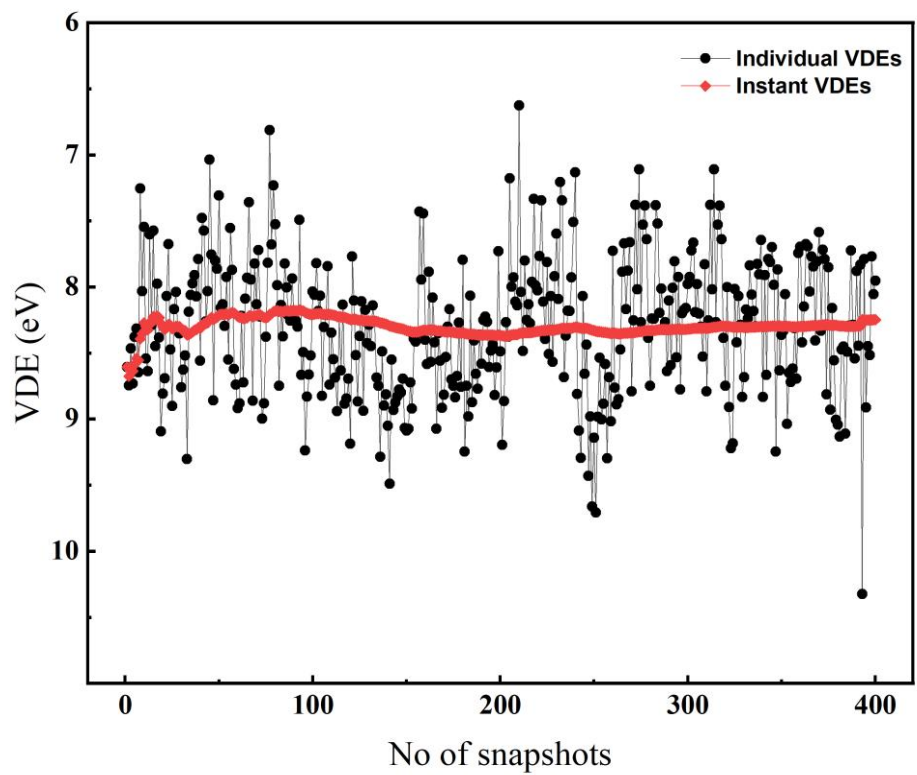

*Figure 6: Vertical detachment energy and instantaneous VDE for 400 snapshots in EA-EOM-DLPNO-CCSD/aug-cc-pVDZ based QM/MM method.*

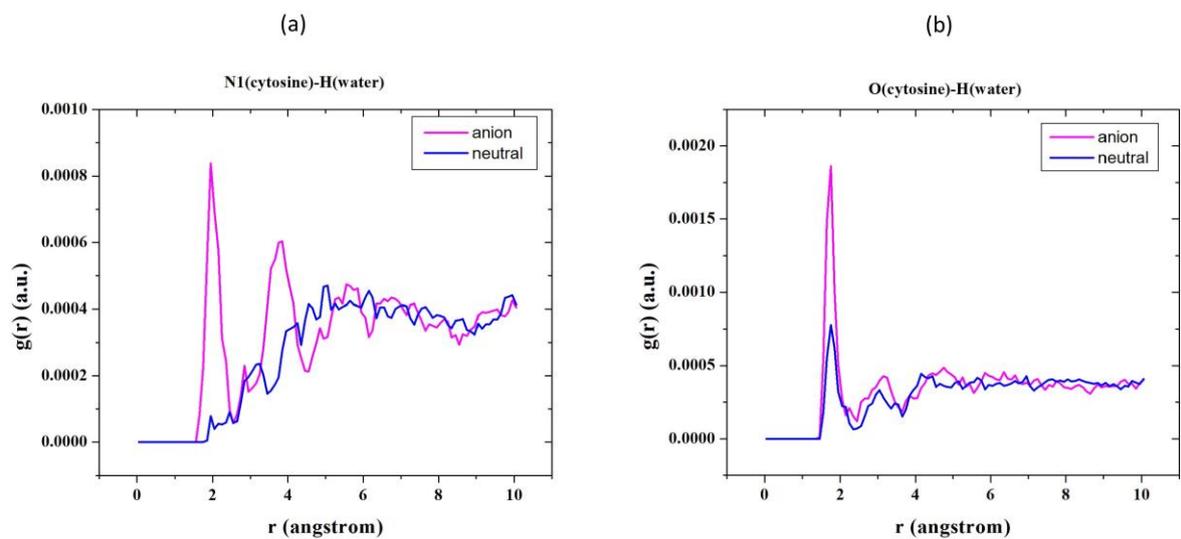

*Figure 7: Radial distribution function plot of (a) primary amine nitrogen of cytosine and hydrogen of water (b) cytosine oxygen and hydrogen of water.*

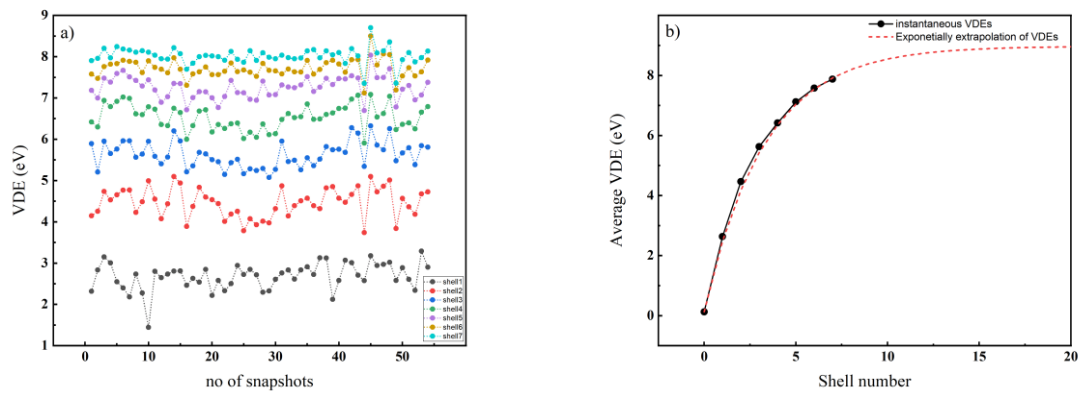

*Figure 8: Shell wise convergence of (a)VDE (b) instantaneous average VDE. A shell length of 2.7 Å has been considered for each water shell.*

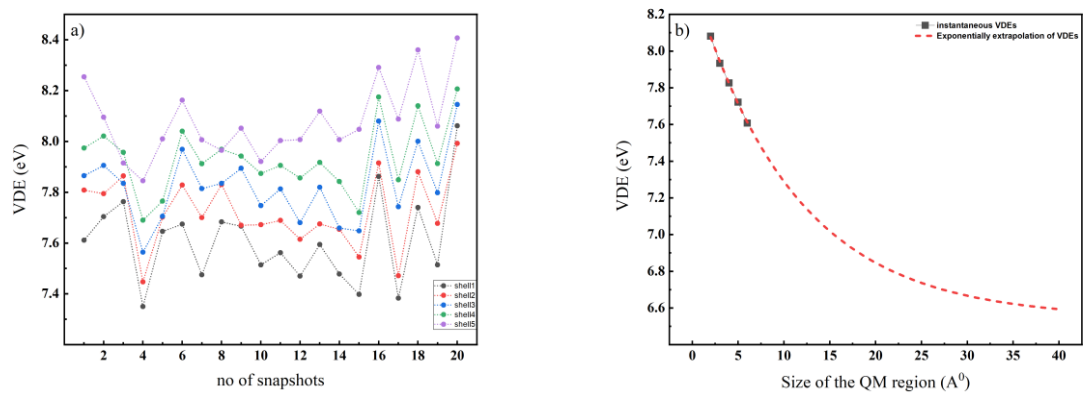

***Figure 9: Change of (a)VDE and (b) instantaneous average VDE in aug-cc-pVDZ basis set with respect to the increase in the size of the QM region.***

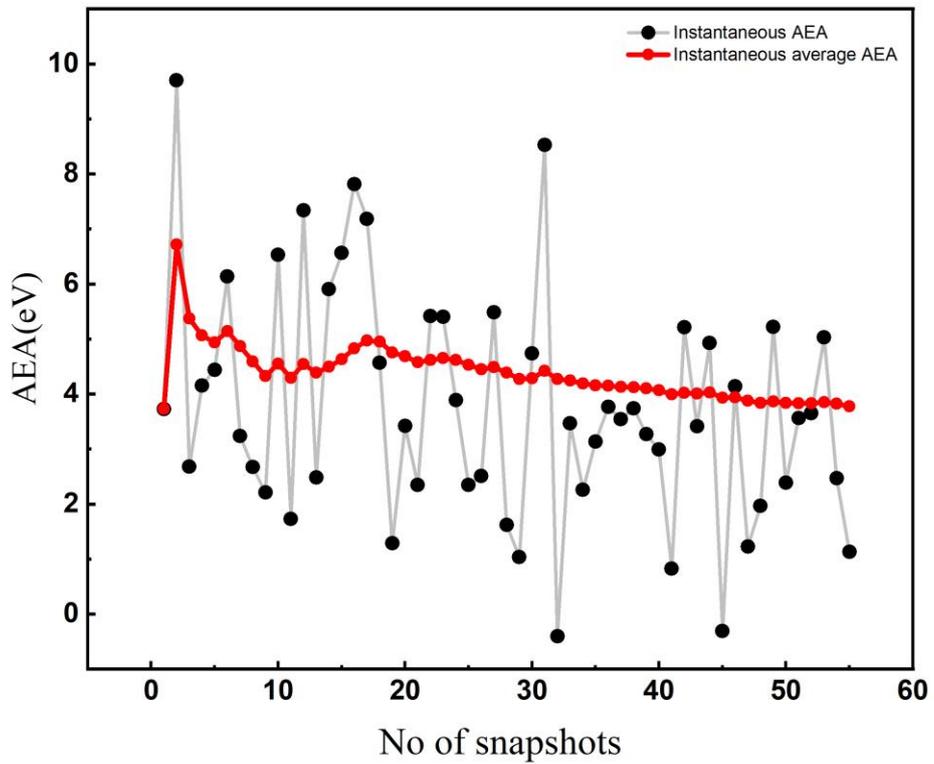

*Figure 10: AEA and instantaneous AEA for bulksolvated cytosine in aug-cc-pVDZ basis set.*

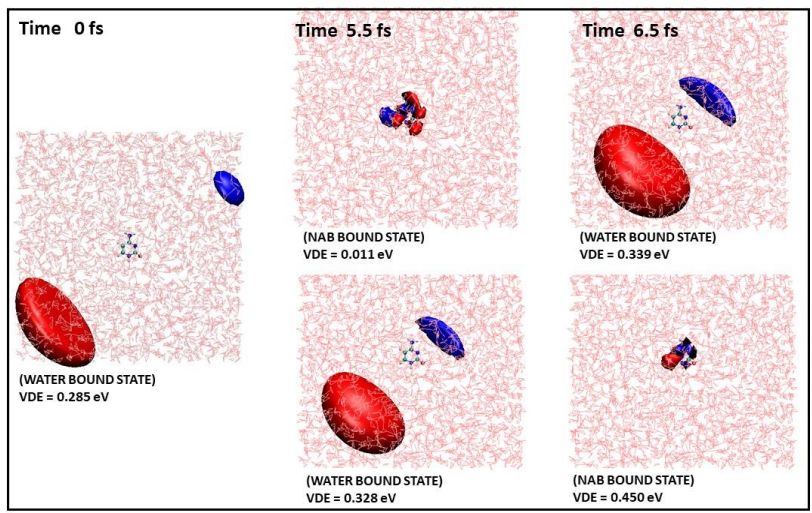

*Figure 11: EA-EOM-DLPNO-CCSD natural orbitals depicting time evolution of anionic state formed by attachment of bulk-solvated electron to cytosine.*

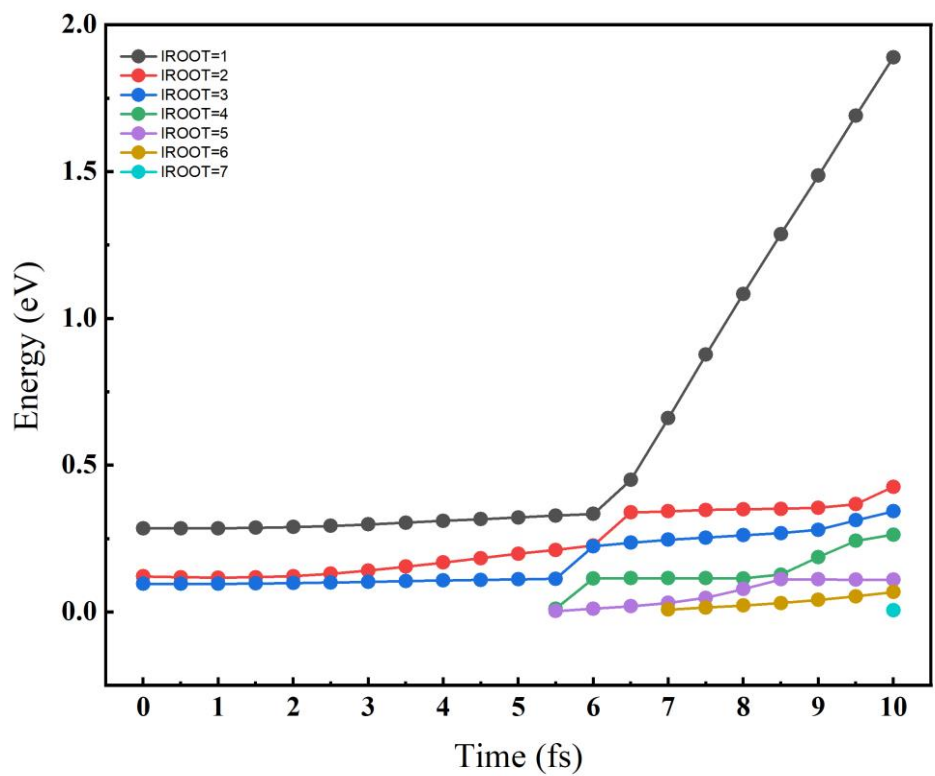

*Figure 12: Time evolution of detachment energy of first six bound anionic state of bulk solvated cytosine anion.*

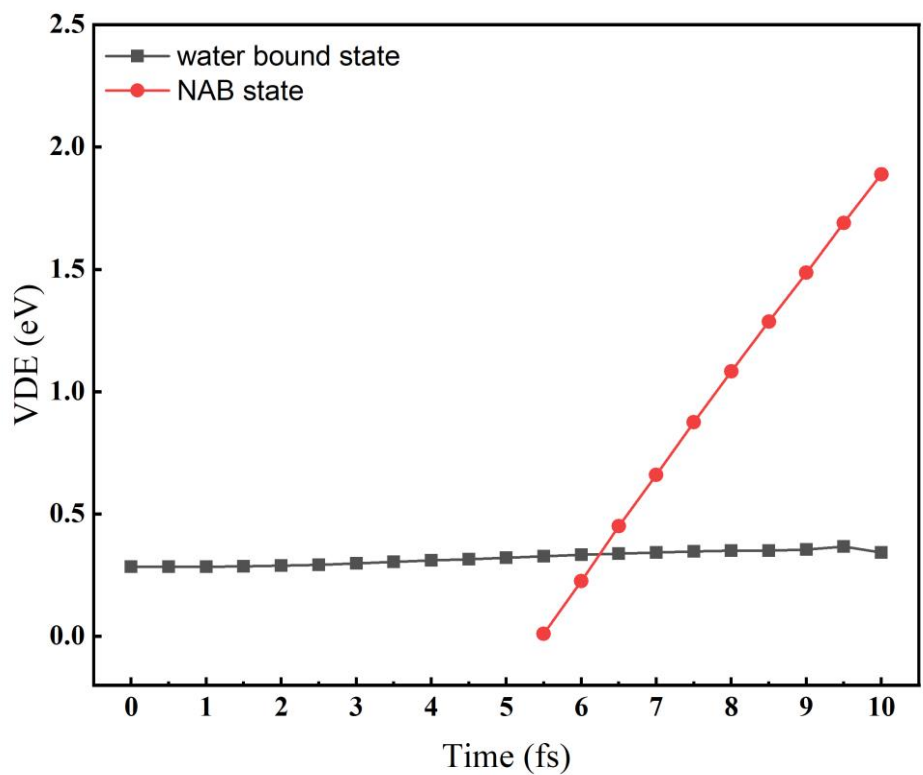

***Figure 13: Time evolution of detachment energy of anionic states on localized water and anionic states localized on cytosine.***